\documentclass[pre,aps,superscriptaddress,showpacs,twocolumn,floatfix]{revtex4-1}

\usepackage{amssymb,amsmath}
\usepackage{graphicx}
\usepackage{nicefrac}
\usepackage{color}
\usepackage{gensymb}
\usepackage{enumerate}

\newcommand{\order}{{\cal O}}

\begin{document}

\title
{3D-2D transition in mode-I fracture microbranching in a perturbed hexagonal close-packed lattice}
%{Microbranching in simulations of mode-I fracture in a three-dimensional perturbed hexagonal close-packed (hcp) lattice}
\author
{Shay I. Heizler}
\email{highzlers@walla.co.il}
\affiliation{Department of Physics, Bar-Ilan University, Ramat-Gan, IL52900 ISRAEL}
\affiliation{Department of Physics, Nuclear Research Center-Negev, P.O. Box 9001, Beer Sheva IL84190, ISRAEL}
\author
{David A. Kessler}
\email{kessler@dave.ph.biu.ac.il}
\affiliation{Department of Physics, Bar-Ilan University, Ramat-Gan, IL52900 ISRAEL}
 

\begin{abstract}

Mode-I fracture exhibits microbranching in the high velocity regime where the simple straight crack is unstable. For velocities below the instability, classic
modeling using linear elasticity is valid. However, showing the existence of the instability and calculating the dynamics post-instability within the linear elastic
framework is difficult and controversial. The experimental results give several indications that the
microbranching phenomenon is basically a three-dimensional phenomenon. Nevertheless, the theoretical effort has been focused mostly in two-dimensional modeling. 
%There some success has been achieved concerning the origin of the instability and the post-instability behavior, particularly within the context of atomistic simulations.
In this work we study the microbranching
instability using three-dimensional atomistic simulations, exploring the difference between the 2D and 3D models. We find that the basic 3D fracture pattern
shares similar behavior with the 2D case. Nevertheless, we exhibit a clear 3D-2D transition as the crack velocity increases, while as long as the microbranches are
sufficiently small, the behavior is pure 3D-behavior, while at large driving, as the size of the microbranches increases, more 2D-like behavior is exhibited.
In addition, in 3D simulations, the quantitative features of the microbranches, separating the regimes of steady-state cracks (mirror) and post-instability (mist-hackle)
are reproduced clearly, consistent with the experimental findings.

\end{abstract}

\maketitle

\section{Introduction}
\label{intro}
Over the last decades, the dynamic instability in mode-I fracture has been extensively studied~\cite{review}. These findings deviate from the two-dimensional classic model for mode-I fracture
with a single crack that propagates in the midline of the sample, based on linear elasticity fracture mechanics (LEFM)~\cite{freund}. This classic
theory, which lacked a supplemental criteria for instability, predicts that a single crack will accelerate to a terminal velocity, which for mode-I fracture is the
Rayleigh surface wave speed, $c_R$. In fact, as long as a single crack
does exist, the crack obeys LEFM predictions~\cite{review_mid,review_new}. However, the experiments
find that at a much lower velocity ($\approx 0.36-0.42c_R$, for a short review, see for example~\cite{adda_bedia}), a dynamic instability occurs, and small microbranches start to appear nearby the
main crack~\cite{marder_jay1,fineberg_sharon1,fineberg_sharon2,fineberg_sharon3,fineberg_sharon5}. The additional energy that has to be spent in creating the new surfaces prevents the crack from accelerating
to the theoretical terminal velocity. LEFM-based universal criteria for branching~\cite{yoffe,eshelby} fail to describe the instability, predicting a much higher critical velocity than in reality.
Moreover, when the small microbranches appear at $v\geqslant v_{cr}$, they present a clear 3D nature. However, when enlarging the driving displacement, the small microbranches reunite,
creating 2D patterns (right before macro-branches appear), especially in PMMA~\cite{fineberg_sharon1,fineberg_sharon2,fineberg_sharon3,fineberg_sharon5}.
\begin{figure}
\centering{
 \includegraphics*[width=7.5cm]{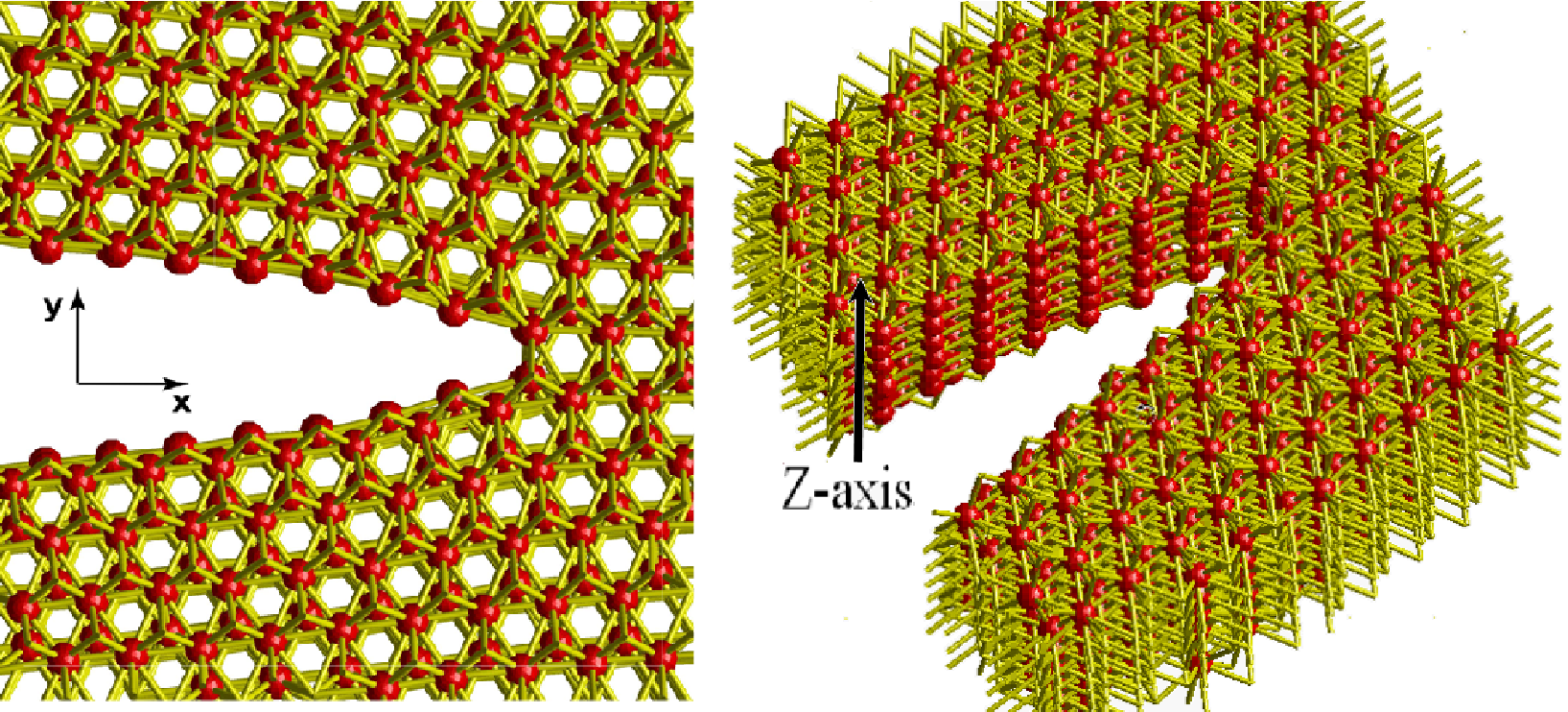}
}
\caption{(color online) A snapshot of the (same) crack tip in steady-state crack using a perturbed hexagonal close-packed (hcp) lattice, from different viewing angles.
Each atom shares 12 nearest neighbors, defining
``bonds" that connect each other by a force law, and is allowed to move in all three coordinates. The crack creates mirror-like pattern. The left snapshot is a clear
XY-plan view while the right has a slight tilt, showing how deep the system is.}
\label{3d_pattern}
\end{figure} 

Lattice models reproduce the existence of steady state cracks~\cite{slepyan,slepyan2}, and via a standard linear stability analysis, they predict the existence
of a critical velocity, when the steady-state cracks becomes linearly unstable~\cite{marderliu,kess_lev3,pechenik,shay1,shay2}. This critical velocity is found to be strongly dependent on
the details of the inter-atomic potential, such as the degree of smoothness of the potential (as it drops to zero), or the amount of dissipation. Simple simulations that use these same
potentials succeed in reproducing the steady-state regime, yielding the exact point of instability, and in reproducing the lattice models results, but fail to describe the 
behavior in the post-instability regime~\cite{fineberg_mar,shay1}. The early efforts on using a binary-alloy model for modeling brittle amorphous materials failed to achieve
steady-state cracks at all~\cite{falk}, although more recent attempts have succeeded in yielding propagating cracks~\cite{procaccia1,procaccia2}.

Recent studies using Zachariasen's~\cite{zachariasen} 2D continuous random network model (CRN) of amorphous materials, a model that also has recently received experimental support
from direct imaging of 2D silica glasses~\cite{tem1}, were used in describing the microbranching instability~\cite{shay3} (using $\order\left(10^4\right)$ 2D particle mesh).
The simulations reproduced qualitatively both the regime of steady-state propagating cracks and the fracture patterns of the microbranches. In addition, using perturbed lattice
models, generated by adding a small amount of disorder to the bond lengths, supplemented by an additional 3-body force-law which penalize rotation of the
bonds away from the natural directions of the lattice, produces similar results~\cite{shay4}. Larger scale simulations ($\order\left(10^6\right)$ particles) using GPU computing 
yields various qualitative and quantitative results of post-instability behavior such as a sharp transition between the regime of steady-state and microbranching, the increase of the derivative
of the electrical resistance across the crack with respect to time (which correlates experimentally with the crack velocity), the correct branching angle and also the power-law behavior of the
branch shapes~\cite{shay5}. All of the theoretical models that were mentioned above employed a 2D description of the problem.

The large scale simulations allow us for the first time to perform three-dimensional (3D) simulations, attacking the microbranching phenomenon which is, at its heart, a
3D phenomenon~\cite{review,fineberg_sharon2,kolvin}, by taking the $\order\left(10^4\right)$ particle mesh and adding a third dimension with $N_Z\approx100$.
The two basic questions that we address using our 3D simulations are:
\begin{enumerate}[(i)]
\item Checking the reliability of the previous 2D simulations, investigating how well the 2D description reproduces the behavior of the more realistic 3D models;
\item Studying for the first time the direct 3D experimental features of the microbranches, which have not previously been modeled.
\end{enumerate}
We note that several 3D fracture molecular-dynamics simulations, containing large numbers of atoms, have been studied previously using different potentials (for
example, see~\cite{abraham1,abraham2,review3d}),
but intensive study concerning the features of the 3D instability and the features of the microbranches have not yet been studied. It is important to note that atomistic simulations
cannot reproduce the fracture patterns on the real physical length scales, of the experiments. However, they try to reproduce {\em scaled} results and scaled structures of the real
fracture length scales. 

\section{Model and General Methodology}
\label{model}
Our simulations consist of $\approx3\cdot10^6$ atoms, which include $1.7\cdot10^7$ bonds (central force-laws), and $\approx3.4\cdot10^7$ 3-body interactions
(see appendix \ref{app_a2} for the exact 
parameter of the 3-body potential that was used). These simulations can be performed in reasonable run times by using parallel GPU computing (see appendix \ref{app_a3}).
\begin{figure}
\centering{
\includegraphics*[width=7.5cm]{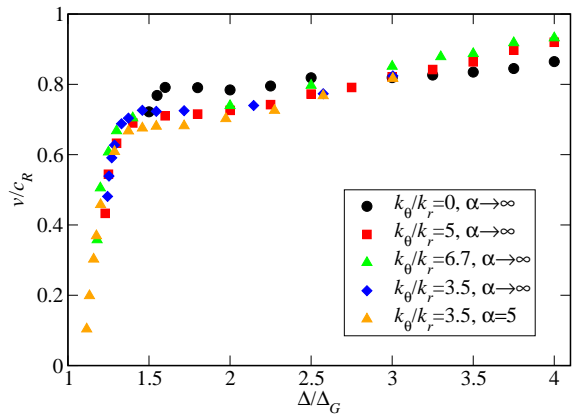}
}
\caption{(Color online) The $v(\Delta)$ curve of the perturbed hcp with different values of 2-body and 3-body force laws. With a finite value of $\alpha$ and
$k_{\theta}/k_r$, the velocity gap shrinks dramatically, yielding the correct experimental behavior.}
\label{quant1}
\end{figure}

We used a perturbed hexagonal close-packed (hcp) structure, which is a 3D extension of the 2D perturbed hexagonal lattice that was studied in~\cite{shay4,shay5}.
As in our 2D studies, the interactions are taken to be only between nearest-neighbors in the unperturbed hcp lattice, with an in-plane lattice constant of $a=4$ and $c=\sqrt{8/3}a$ (see
Fig. \ref{hcp1} in appendix \ref{app_a1}).
Every atom has 12 closest neighbors. We add a small amount of
disorder to the bond lengths, $a_{i,j}=(1+\epsilon_{i,j})a$
where $\epsilon_{i,j}\in[-b,b]$, and $b$ is constant, and in this work is set to $b=0.1$ (for the system shape, see Fig. \ref{hcp2} in appendix \ref{app_a2}). In most of our simulations,
we employed a piecewise-linear radial force law (in this work, $k_r=1$) between the initially neighboring atoms. However, in some of them we used a more physical smooth force law,
using a smoothness parameter $\alpha$, which when $\alpha\to\infty$ reproduces the piecewise-linear model (see appendix \ref{app_a2}).
In addition, we add a 3-body potential and Kelvin type viscosity, as described in detail in our 2D lattice studies~\cite{shay4,shay5}.
We relax the system, and then we strain the lattice under a mode-I tensile loading with a given constant strain grip boundary condition corresponding to
a given driving displacement $\pm\Delta$ (which is normalized relative to the Griffith displacement $\Delta_G$) of the edges and seed the system with an initial crack.
For a detailed discussion regarding the model and the governing equations, see appendixes \ref{app_a1} and\ref{app_a2}.
The crack then propagates via the same molecular dynamics Euler scheme (the simulations were always stable using a reasonable value of $dt$, so we have not needed any
more sophisticated numerical schemes). In Fig. \ref{3d_pattern} we present close-in snapshots of the (same) crack tip in a steady-state
crack from different viewing angles. We can see that at low driving displacement the crack is actually 2D in nature. 

\section{Microbranching instability in 3D-Perturbed Lattice}

The crack velocity $v$ (which we normalize to the Rayleigh wave speed $c_R$) increases with $\Delta/\Delta_G$ (see Fig. \ref{quant1}). We define the Rayleigh wave speed
here as that calculated from $c_l$ and $c_t$ (the longitude and the transverse wave speeds) in the XY-plane ($(0001)$ in the crystallographic notation), which is the major fracture surface
in our simulations (there is a symmetry along the Z-axes in steady-state cracks, see appendix \ref{app_b}). We can see that using a perfect non-perturbed
lattice (in these simulations we used also $k_{\theta}=0$, in addition to $b=0$, but this result is valid for all value of $k_{\theta}$),
we get a (non-physical) velocity gap (like in 2D~\cite{marderliu,kess_lev3,pechenik,shay1}), in which slow cracks are prohibited.
However, adding disorder and the 3-body force-law, the velocity gap shrinks, and by using a finite value of $\alpha$, the velocity
gap shrinks dramatically with steady-state cracks in almost zero velocities, yielding the correct experimental behavior~\cite{review}.
\begin{figure}
\centering{
 \includegraphics*[width=7.5cm]{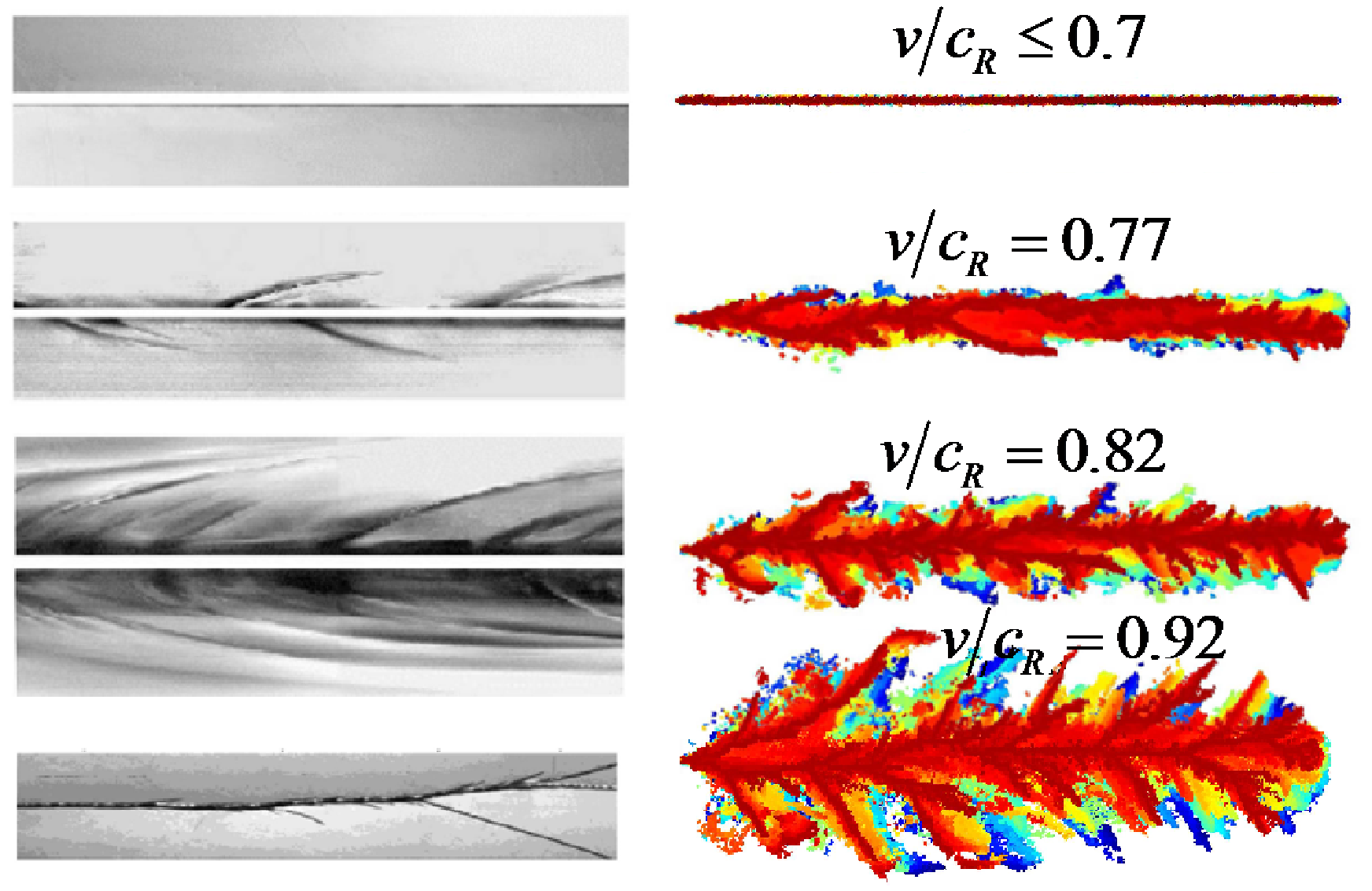}
}
\caption{(Color online) A XY plan view of the experimental microbranching phenomenon in PMMA taken from~\cite{fineberg_sharon2,fineberg_sharon5} (left) for increasing driving displacement.
In color (right) we see our simulations XY plan view, where the color denotes the Z-location of the broken bonds (dark red for top edge and dark blue for bottom edge).
We can see that despite the quite noisy simulations, in general the qualitative picture is quite good. The upper picture yields a mirror-like steady-state crack and is valid for
all $v/c_R\leqslant 0.7$.}
\label{XY}
\end{figure}

\begin{figure}
\centering{
\includegraphics*[width=7.5cm]{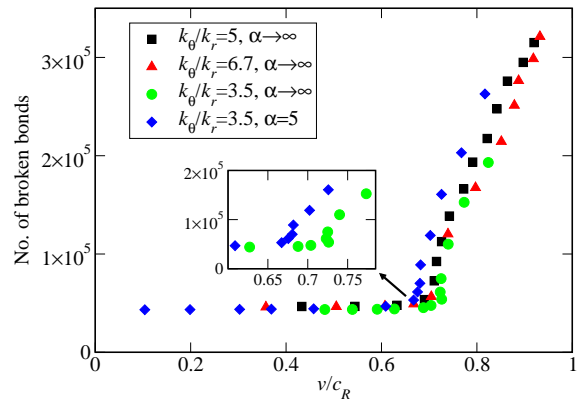}
}
\caption{(Color online) The total number of broken bonds as a function of the crack velocity (the constant number of bonds in the small velocity regime represents
the broken bonds of the main crack). A clear transition between the steady-state regime and microbranching behavior can be seen. In the small box there is a zoomed
picture of the transition area.}
\label{quant2}
\end{figure}

\begin{figure}
\centering{
\includegraphics*[width=7.5cm]{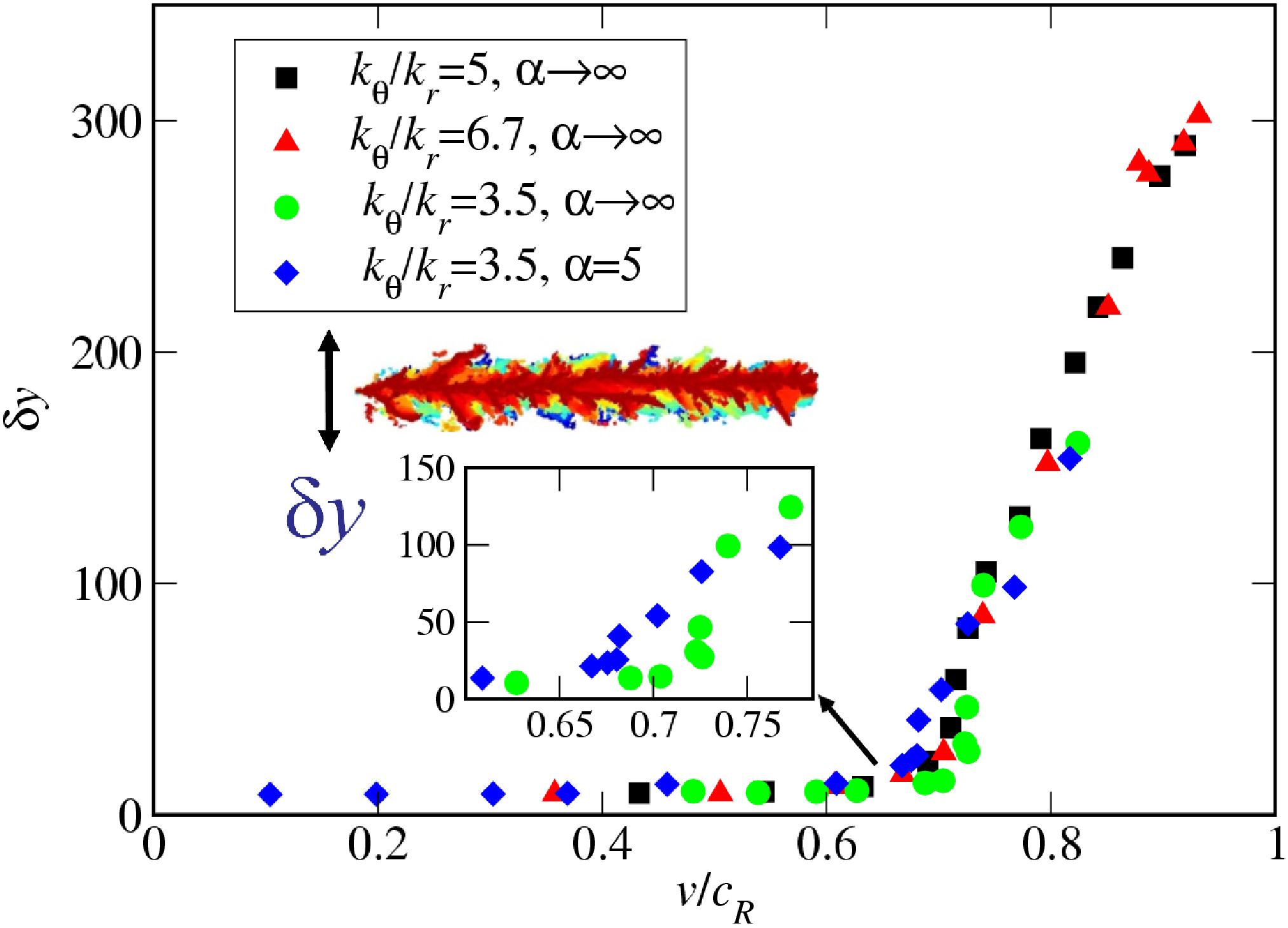}
}
\caption{(Color online) The width of the fracture region as a function of the crack velocity. A clear transition between the steady-state regime and microbranching behavior can be seen.
In the small box there is a zoomed-in picture of the transition area.}
\label{quant3}
\end{figure}

In. Fig. \ref{XY} we show several microbranching patterns (top views), both experimental (in PMMA) and from our 3D
simulations using $k_{\theta}/k_r=5$ (where the color denotes the Z-location).
The broken bonds are plotted in the fractured system, and their Z-location can be associated with the color, where dark red represents the top edge and dark blue for bottom edge.
We see that below the critical velocity, in
the regime of steady-state cracks, the crack has a ``mirror" surface.
Increasing the driving displacement,
small microbranches appear nearby the main crack, while the size of the microbranches increases dramatically with the driving, yielding at first a ``mist" surface and with large
$\Delta$, a ``hackle" surface. Despite the noisy results (due to the relatively small size of the simulations),
the  pictures are qualitatively quite similar to the experimental findings, at least in the sense that the microbranches increase dramatically with the driving displacement, yielding
eventually large macro-branches (in the simulations, a ``macro-branch" is a branch that reaches the end of the sample, like in the experiments, on a different length scale). A quantitative
(scaled) overview is presented in Figs. \ref{quant2}-\ref{quant3}.
We note that without a 3-body force-law we do not get the microbranching pattern, but rather a cleavage-like
behavior (with or without the presence of disorder). Using too strong a 3-body force law
($k_{\theta}/k_r=6.7$), yields microbranches that propagate in straight lines with the natural angle of the lattice ($60\degree$), which is again non-physical.

The transition between the regime of steady-state cracks and the post-instability side-branching regime is very sharp in the 3D simulations. In Figs. \ref{quant2}-\ref{quant3} we present
two quantitative parameters that demonstrate this sharp transition (in the small box there is a zoomed picture of the transition area). In Fig. \ref{quant2} the total number
of broken bonds as a function of the crack velocity is displayed. In the small
velocity regime, only the bonds necessary for yielding a single main crack are broken. Beyond the critical velocity, the number of broken bonds increases linearly,
as in the experiments~\cite{fineberg_sharon2}, and broadly similar (although much sharper here) to what is seen in the hexagonal perturbed 2D lattice~\cite{shay5}. 

In Fig. \ref{quant3} we measure $\delta y$, the width of the microbranching region, as a function of the crack velocity (see definition inside Fig. \ref{quant3}).
$\delta y$ is a second measure of the size of the microbranches.
As above, a sharp transition can be seen between the single crack and microbranching regimes. We note that using the piecewise-linear force law, the critical velocity $v_{\mathrm{cr}}$
seems to be very close to the Yoffe criterion (which is $\approx0.73c_R$). But, as we showed previously in 2D, the quantitative value of the $v_{\mathrm{cr}}$ can be controlled via the inter-atomic potential
parameters, such as $\alpha$ and $\eta$ (see appendix \ref{app_a2} for explicit definitions of these parameters)~\cite{kess_lev3,shay1,shay5}.
We can see that using a finite value of $\alpha$, the critical velocity decreases (see the small boxes in Figs. \ref{quant2}-\ref{quant3} for a given $k_{\theta}$),
to the exact value of the 2D simulations; in $\alpha\to\infty$ we reproduce the 2D critical velocity $v_{\mathrm{cr}}\approx0.73c_R$ (see Fig. 7 in~\cite{pechenik}), while also with $\alpha=5$
we reproduce the 2D value, $v_{\mathrm{cr}}\approx0.68c_R$ (see Fig. 4(a) in~\cite{shay1}). That means that the critical velocity is not universal and is potential-dependent.
Thus, for example we can
vary the values of $\alpha$ and $\eta$ to reproduce the exact experimental critical velocity of a given material, very much like we did in 2D~\cite{shay1,shay2}.
In both Fig. \ref{quant2} and \ref{quant3}, the results appear insensitive to the exact value of $k_{\theta}$, despite the fact that the microbranches in the two cases appear different.

\begin{figure}
\centering{
 \includegraphics*[width=7.5cm]{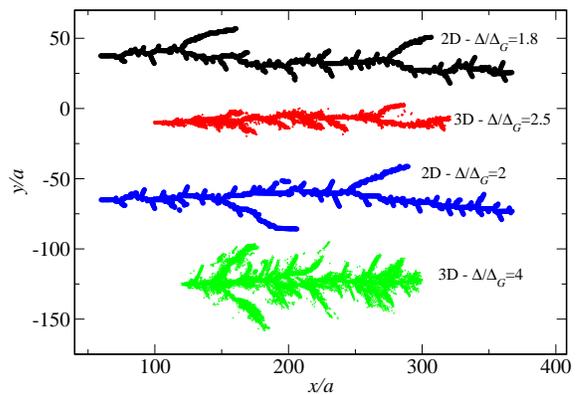}
}
\caption{(Color online) A comparison between the fracture pattern (in lattice scale units) of a pure 2D perturbed hexagonal simulation and thin
slices of the 3D fracture pattern using our perturbed hcp lattice. Despite the greater noise of the 3D simulations, the patterns look quite similar.
The 2D patterns are taken from~\cite{shay4}.}
\label{2d3d}
\end{figure}

In addition, we can cut thin horizontal slices from the XY fracture pattern, yielding 2D patterns and compare them to pure 2D fracture patterns~\cite{shay4}.
In Fig. \ref{2d3d} we present two fracture patterns of a 2D perturbed hexagonal
lattice and and two 2D slices of the 3D hcp perturbed lattice, one for relatively small driving and one for large driving displacement. We can see despite the
relatively large noise (resulting from the breaking of one or few bonds) that characterizes the 3D simulations, the patterns are quite similar to pure 2D simulations
for small driving displacement.
This fact is encouraging and supports the assumption that for at least some features (e.g., XY-plane features of the microbranches), the 2D studies are relevant.
However, for large driving displacement, the 3D patterns look rather different from the 2D patterns, though the fracture-pattern is still much more developed
at large driving displacements. Nevertheless, we note that different horizontal slices of the same 3D fracture pattern (for different Z)
yield different patterns. This fact indicates that for the 3D regime, as
long as the microbranches are sufficiently small, there is no symmetry along the Z-axis. Note that the driving displacement $v/c_R$ required
to produce a given amount of side-branching is much greater than in 2D, since out of plane bonds are being broken as well.

\section{The 3D-2D transition }

Moreover, we can compare our 3D simulations to the 3D experimental properties of the microbranches. Experimental post-mortem pictures of the XZ-plane of the fractured surface
by Sharon \& Fineberg~\cite{fineberg_sharon2} reveal that nearby the origin of instability, the microbranches are localized in the Z-axis. At large velocities, the microbranches merge, creating
a Z-plane quasi-symmetric pattern, yielding a 3D-2D transition~\cite{review,fineberg_sharon2,fineberg_sharon3,fineberg_sharon5}. In PMMA (as opposed to glasses or gels),
nice symmetric ``2D"-like strips are created in association with the largest microbranches~\cite{fineberg_sharon2}.
\begin{figure}
\centering{
 \includegraphics*[width=7.5cm]{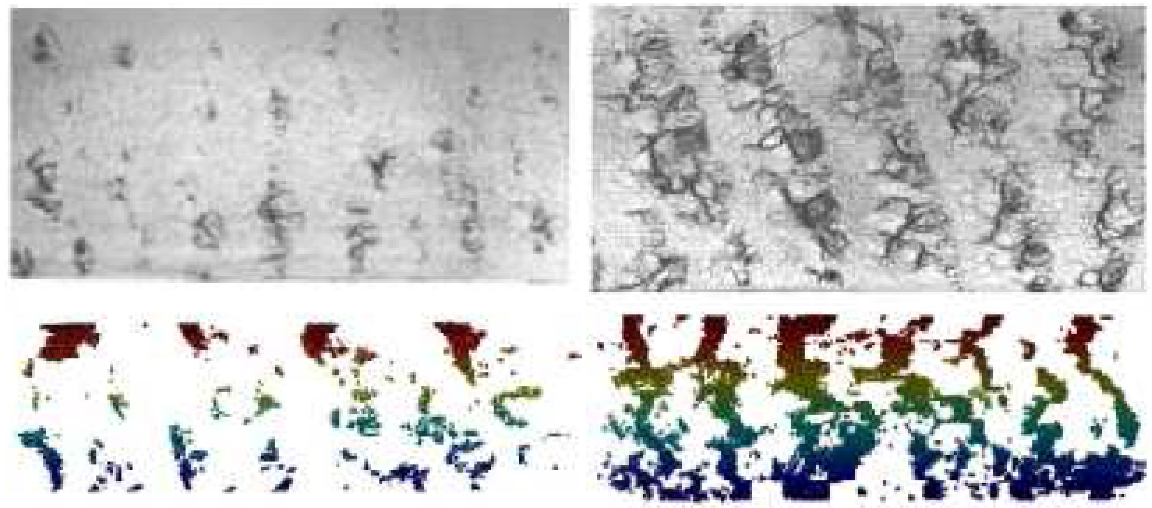}
}
\caption{(Color online) Top row: A XZ-plane view of the experimental microbranching phenomenon in PMMA of Sharon \& Fineberg experiments (which are taken from~\cite{fineberg_sharon2}), along
with simulations results (where the color denotes the Z-location of the broken bonds for presentation reasons),
for small driving on the left, and for large driving displacement on the right. Bottom row: The simulational XZ plane view. We see that
very much like the experiments, at small driving displacement the microbranching is ``3D" and for large driving displacement, the microbranches are ``2D" in character.}
\label{XZ}
\end{figure}

In Fig. \ref{XZ} we present two experimental pictures of XZ-plane of the fracture surface that demonstrates the 2D-3D transition in PMMA, taken from Ref.~\cite{fineberg_sharon2}. Below, we
depict XZ slices taken at a constant distance from the main crack (relative to the Y-axis) of our 3D simulations (the pictures from the main crack plane itself are too noisy, due to our finite
size simulations). We see that the fracture patterns looks surprisingly similar. At small driving displacement ($\Delta/\Delta_G=2.5$ in the simulations),
right beyond the critical velocity, the microbranches are localized in the Z-directions, yielding purely 3D behavior in both the experiments and the simulations.
Increasing the driving displacement further ($\Delta/\Delta_G=4$ in the simulations), the microbranching increases in the Z-direction from top to bottom of the sample, yielding
a 2D type behavior. The periodic stripes structure is a result of the periodic microbranches in the XY-plane (Fig. \ref{XY})~\cite{fineberg_sharon1}. After the onset of branching,
the energy flowing into the crack tip is divided between the main crack and the daughter cracks.
The daughter cracks, which compete with the main crack, have a finite (similar) lifetime, because the main crack can outrun them and screen the daughter cracks from the surrounding 
stress field. The daughter cracks then die and the energy that had been diverted from the main crack returns. The scenario then repeats itself,
causing the branching pattern to be more or less, periodic.

As a matter of fact, these large microbranches result from the merging of several small microbranches, as we can see carefully in Fig. \ref{XZ} (there is not a perfect
symmetry along the Z-axis; for different Z, the microbranch propagates different distances). This behavior shares similar features with recent experimental work~\cite{kolvin}.

\begin{figure}
\centering{
 \includegraphics*[width=7.5cm]{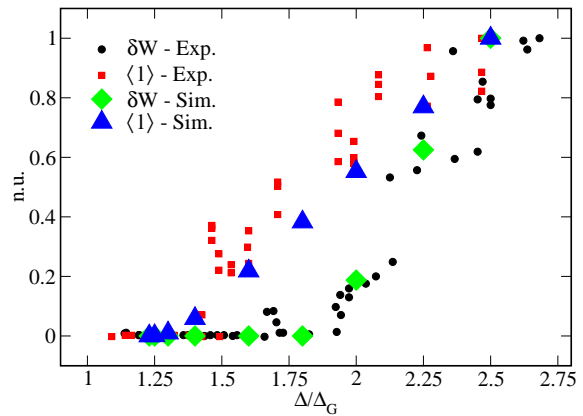}
}
\caption{(Color online) A comparison between experimental and simulations results for the average length of microbranch and microbranch
width (normalized to the maximal yielded value) as a function of the driving displacement (normalized to the Griffith value).
The experimental data is taken from~\cite{fineberg_sharon2,review}.}
\label{3d_branch}
\end{figure}

We can now quantify this 2D-3D transition (of course in normalized units). Looking carefully in the PMMA experimental results, we can see that the region of
instability, $v=v_{cr}\approx 340 \mathrm{m/s}$ (Fig. 11(a) in~\cite{fineberg_sharon2}) is quite differ from the point of 2D-3D transition,
$v\approx 550 \mathrm{m/s}$ (Fig. 19 in~\cite{fineberg_sharon2}), ensuring the fact that at first (near $v\approx v_{cr}$) the microbranches are ``3D", while only for larger velocities,
they become ``2D".
In Fig. \ref{3d_branch}, we plot the width of the largest microbranch (in the Z-direction) in the 3D simulations for a given $\Delta/\Delta_G$, along with the total number
of broken bonds (from Fig. \ref{quant2}), both of them are normalized to their largest value. We plot them both as a function of $\Delta/\Delta_G$ and not as a function of $v/c_R$,
since the crack velocities are an output parameter (and in our simulations are much higher than the PMMA experimental results). For the experimental results, we used Fig. 17 in~\cite{review} for transferring the data from $v/c_R$ to $\Delta/\Delta_G$.

We can see that the 3D simulations results are reproducing the 2D-3D transition almost perfectly. At $\Delta/\Delta_G\approx 1.3$, in both the experimental and the
simulations results, small microbranches start to appear the main crack. Those microbranches are localized in the Z-direction, while only at $\Delta/\Delta_G\approx 1.8-1.9$, the width of the microbranches increases dramatically,
yielding ``2D-microbranches", when several microbranches reunite, covering the whole Z-directions,
yielding a 3D-2D transition.

\section{Summary and Future work}

In conclusion, as long as we look in the XY-plane, the 3D simulations share similar features as the 2D simulations, and quantitative measures as to the total number of microbranches
or the size of the opening of the microbranches as a function of crack velocity look the same. On the other hand, our current simulations also reproduce pure 3D features,
especially the XZ-plane patterns, when the 3D-2D transition occurs. Thus, we believe that the lattice models and simulations offer a good theoretical framework for studying the microbranching
instability, including the 3D effects. 
We are left with the following question. In 2D~\cite{shay5}, enlarging the system allows quantitative study
of the branches. How will behave the 3D system on a larger scale? The answers should be attainable within the scope of available supercomputers, using thousands of nodes, or tens of GPU's. 

\appendix

\section{Generating the perturbed lattice}
\label{app_a1}
\begin{figure}
\centering{
 \includegraphics*[width=5cm]{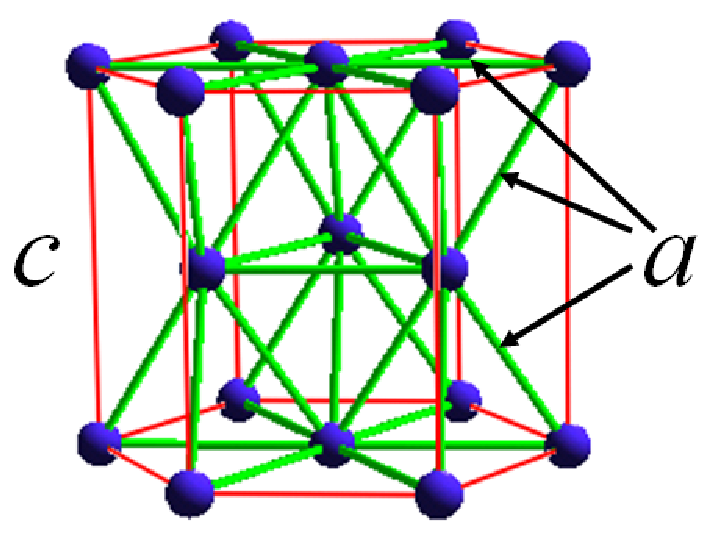}
}
\caption{(color online) A diagram of the unit cell of the ideal hexagonal closed packed (hcp) lattice, where $c=\sqrt{8/3}a$. Each atom has 12 nearest neighbors, 6 in the
XY-plane (reproducing the 2D hexagonal lattice).}
\label{hcp1}
\end{figure}

We start with a perfect ideal hexagonal close-packed (hcp) lattice, where $c=\sqrt{8/3}a$ (see Fig. \ref{hcp1}) when each atom has 12 nearest neighbors, 6 in the XY-plane,
(yielding a 2D hexagonal lattice) and 6 in the Z-direction (3 up and 3 down).
As in the 2D-studies~\cite{shay4,shay5}, we randomize the length of each ``bond", $a_{i,j}$:
\begin{equation}
a_{i,j}=(1+\epsilon_{i,j})a,\qquad i=1,2,\dots,n_{\mathrm{atoms}}, j\in {\cal N}(i)
\label{perturbed}
\end{equation}
where $\epsilon_{ij}\in[-b,b]$, and $b$ is constant for a given lattice. In this work we set $b=0.1$ and $a=4$. ${\cal N}(i)$ refers
to the nearest-neighbors of site $i$. 

\section{The equations of motion}
\label{app_a2}
In most of our calculations, between each two atoms there is a piecewise linear radial force (2-body force law) of the form:
\begin{equation}
\vec{f}^r_{i,j}=k_rk'_{i,j}(\vert\vec{r}_{i,j}\vert-a_{i,j})\hat{r}_{j,i},
\label{force_Radial}
\end{equation}
where:
\begin{equation}
k'_{i,j}\equiv\theta_H\left(\varepsilon-\vert\vec{r}_{i,j}\vert\right).
\label{linear}
\end{equation}
The Heaviside step function $\theta_H$ guarantees that the force drops immediately to zero when the distance between two atoms $\vert\vec{r}_{i,j}\vert$ reaches
a certain value $\varepsilon>a_{i,j}$ (the breaking of a ``bond").
In this work we set $\varepsilon=a+1$. Alternatively to Eq. \ref{linear}, we can use a smoother force law,
which instead of a sharp failure at $\vert\vec{r}_{i,j}\vert=\varepsilon$, has a more realistic smooth transition wherein the force law drops to zero of the form~\cite{kess_lev3,shay1}:
\begin{equation}
k'_{i,j}\equiv\frac{1+\tanh [\alpha (\varepsilon -\vec{r}_{i,j})]}{1+\tanh (\alpha)}
\label{nonlinear}
\end{equation}
where $\alpha$ is the smoothness parameter, such that when $\alpha\to\infty$ the force law reverts to the piecewise linear force law. The results in this paper
refer to the piecewise linear model, unless mentioned otherwise.

In addition there is a 3-body force law that depends on the cosine of the angles between each set of 3 neighboring atoms, defined of course by:
\begin{equation}
\cos\theta_{i,j,k}=\frac{\vec{r}_{i,j}\cdot\vec{r}_{i,k}}{\vert\vec{r}_{i,j}\vert\vert\vec{r}_{i,k}\vert},
\label{cos_teta}
\end{equation}
that acts on the central atom (atom $i$) of each angle, and may be expressed as:
\begin{align}
\label{force_teta}
& \vec{f}^{\theta}_{i,(j,k)}=k_{\theta}(\cos\theta_{i,j,k}-\cos\theta_C)\frac{\partial\cos\theta_{i,j,k}}{\partial\vec{r}_i}\,
k'_{i,j}k'_{i,k}\hat{r}_i=\\
& k_{\theta}(\cos\theta_{i,j,k}-\cos\theta_C)\left[\frac{\vec{r}_{i,j}+\vec{r}_{i,k}}{\vert\vec{r}_{i,j}\vert\vert\vec{r}_{i,k}\vert}+ \nonumber
\frac{\vec{r}_{j,i}(\vec{r}_{i,j}\cdot\vec{r}_{i,k})}{\vert\vec{r}_{i,j}\vert^3\vert\vec{r}_{i,k}\vert}+\right.\\
&\left.\frac{\vec{r}_{k,i}(\vec{r}_{i,j}\cdot\vec{r}_{i,k})}{\vert\vec{r}_{i,j}\vert\vert\vec{r}_{i,k}\vert^3} \right]k'_{i,j}k'_{i,k}, \nonumber
\end{align}
while the force that is applied on the other two atoms (atoms $j,k$) may be expressed as:
\begin{align}
\label{force_teta2}
& \vec{f}^{\theta}_{j,(i,k)}=k_{\theta}(\cos\theta_{i,j,k}-\cos\theta_C)\frac{\partial\cos\theta_{i,j,k}}{\partial\vec{r}_j}
k'_{i,j}k'_{i,k}\hat{r}_j=\\
& k_{\theta}(\cos\theta_{i,j,k}-\cos\theta_C)\left[\frac{\vec{r}_{k,i}}{\vert\vec{r}_{i,j}\vert\vert\vec{r}_{i,k}\vert}+
\frac{\vec{r}_{i,j}(\vec{r}_{i,j}\cdot\vec{r}_{i,k})}{\vert\vec{r}_{i,j}\vert^3\vert\vec{r}_{i,k}\vert} \right]k'_{i,j}k'_{i,k} \nonumber
\end{align}
Of course, the forces satisfy the relation:
$\vec{f}^{\theta}_{i,(j,k)}=-(\vec{f}^{\theta}_{j,(i,k)}+\vec{f}^{\theta}_{k,(i,j)})$.
The 3-body force law drops immediately to zero when using a piecewise linear force law
when the bond breaks (Eq. \ref{linear}), or may be taken to vanish smoothly, using Eq. \ref{nonlinear}.
\begin{figure}
\centering{
 \includegraphics*[width=5cm]{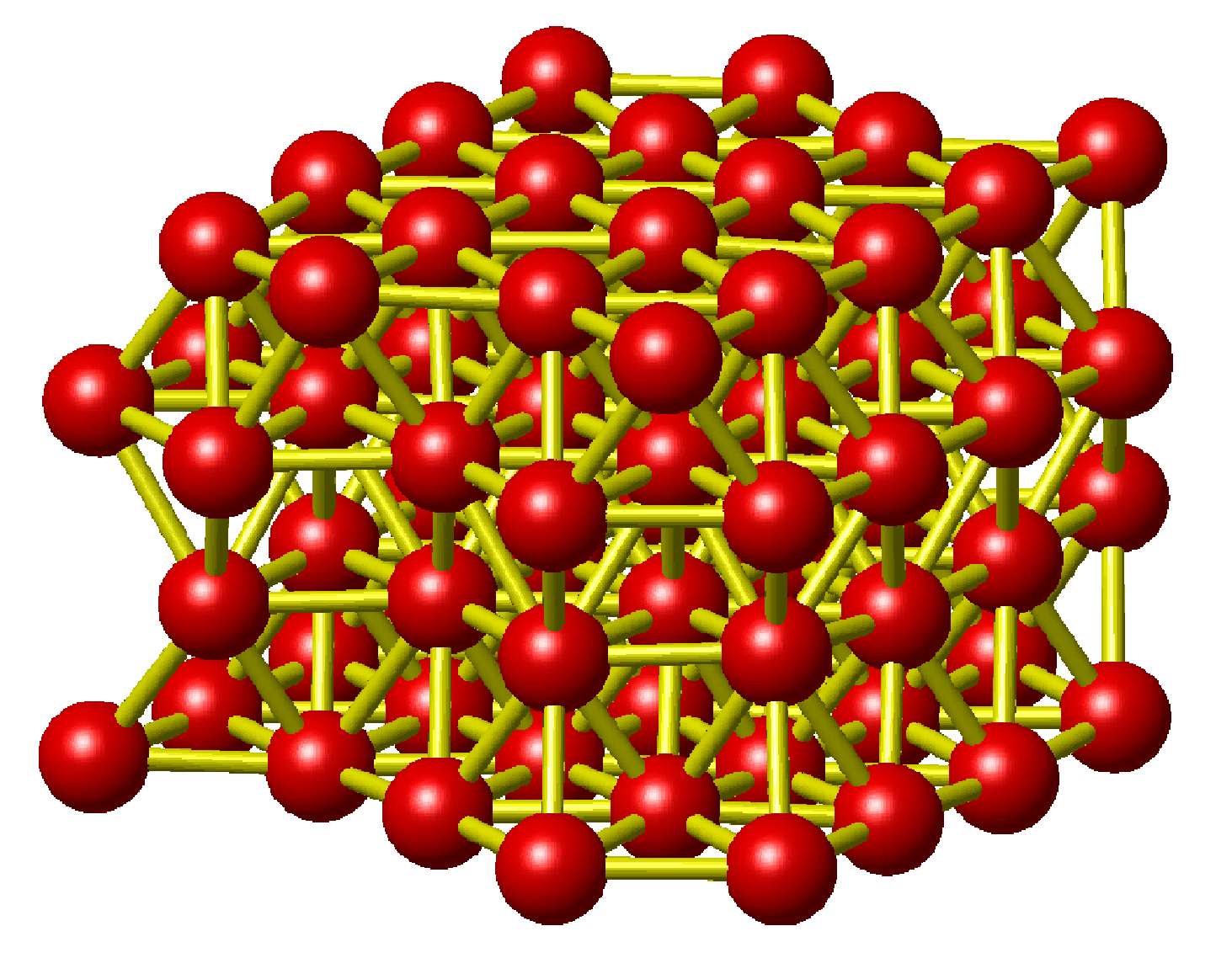}
}
\caption{(color online) A small-scale perturbed hcp yielded by Eq. \ref{perturbed}, after relaxing the system under Eqs. \ref{force_Radial}-\ref{motion_equations}.}
\label{hcp2}
\end{figure} 

We note that in the 3D-case, there are a lot of possible angles between each set of 3 bonds. To shorten the run-times (the calculation of the 3-body force law is extremely time consuming),
in most of our calculations, we do not
include all the possible angles between triplets, but only 12 of them. We chose to take the 6 $60\degree$ angles inside the XY-plane (for reproducing the 2D-hexagonal problem that was studied before
for $N_z=1$), and another 6 angles, three $60\degree$ that connect each atom with its two neighbors that are located in the upper parallel plane, and three angles in the
lower parallel plane (For convenience, see Fig. \ref{hcp1}).
However, in some of our calculations, we used all the 24 $60\degree$ angles, while the results do not vary qualitatively, and the fracture patterns remain similar.
There is a certain preferred angle $\theta_C$ for which the 3-body force law vanishes which is set to $\theta_C=\nicefrac{\pi}{3}$.

In addition, it is convenient to add a small Kelvin-type viscoelastic force proportional to the relative velocity between the two atoms of
the bond $\vec{v}_{i,j}$~\cite{kess_lev1,kess_lev3,pechenik,shay1}:
\begin{equation}
\vec{g}^r_{i,j}=\eta(\vec{v}_{i,j}\cdot\hat{r}_{i,j})\,k'_{i,j}\hat{r}_{i,j},
\label{viscous}
\end{equation}
with $\eta$  the viscosity parameter. The viscous force vanishes after the bond is broken, governed by $k'_{i,j}$. The imposition of a small amount of such a viscosity acts to stabilize the system
and is especially useful in the relatively small systems simulated herein.

The set of equations of motion of each atom is then:
\begin{equation}
m_i\vec{\ddot{a}}_i=\sum_{j\in12\;nn}\left(\vec{f}^r_{i,j}+\vec{g}^r_{i,j}\right)+\sum_{j,k\in12\;nn}\vec{f}^{\theta}_{i,(j,k)}+\sum_{j\in24\;nn}\vec{f}^{\theta}_{j,(i,k)}.
\label{motion_equations}
\end{equation}
In this work the units are chosen so that the radial spring constant $k_r$ and the atoms mass $m_i$ is unity.

After defining the steady-state optimal length of each bond $a_{i,j}$ by Eq. \ref{perturbed}, we first relax the system under the equations of motion, Eqs. \ref{force_Radial}-\ref{motion_equations}
with a small amount of viscosity, yielding the minimal-energy locations of the atoms in the lattice. In Fig. \ref{hcp2}, we can see a small-scale 3D perturbed hcp using our model.

After relaxing the initial lattice, we strain the lattice under a mode-I tensile loading with a given constant strain corresponding a given driving displacement $\pm\Delta$ of the edges
and seed the system with an initial crack. The crack then propagates via the same molecular dynamics Euler scheme using Eqs. \ref{force_Radial}-\ref{motion_equations}.

\section{Parallelization by GPU computing}
\label{app_a3}
As mentioned in Secs. \ref{intro} and \ref{model}, running 3D simulations, using approximately $3\cdot 10^6$ particles cannot reasonably be performed by a singe CPU, and thus force us to use multi-thread computing.
We choose to use GPU computing, parallelizing the code via CUDA~\cite{padon,cooper}, akin to what we implemented before in 2D~\cite{shay5}. This kind of
programing forces the programmer to use the different levels of memory carefully~\cite{cooper},
which makes possible achieving an acceleration up to $\approx 100$ faster than a regular $C$ code using a single CPU. This tool makes possible the simulation of millions
of atoms in  reasonable simulation times. See the appendix of Ref.~\cite{shay5} for more implementation details. In our simulations,
we used $132\cdot310\cdot70\approx3\cdot 10^6$ particles ($N=65$ in the Slepyan model notation).

\section{The Rayleigh surface wave speed for hcp with $k_{\theta}\ne 0$ lattices}
\label{app_b}

Since the models in this paper use a 3-body potential law (aside by the central two-body force law) with $k_{\theta}\ne0$, we need to recalculate the 
Rayleigh wave speed $c_R$, which is the terminal velocity for mode-I fracture for different $k_{\theta}/k_r$. The most convenient way to calculate
the Rayleigh wave speed is to calculate first the longitude (primary) $c_l$ and the transverse (secondary) $c_t$ wave speeds and then, to calculate the
Rayleigh wave speed via the well-known formula~\cite{achenbach}:
\begin{equation}
\left(1-\frac{c_R^2}{c_t^2}\right)^2-4\left(1-\frac{c_R^2}{c_l^2}\right)^{\nicefrac{1}{2}}\left(1-\frac{c_R^2}{c_t^2}\right)^{\nicefrac{1}{2}}=0
\label{cr}
\end{equation}

\begin{figure}
\vspace{5mm}
\centering{
\includegraphics*[width=7.5cm]{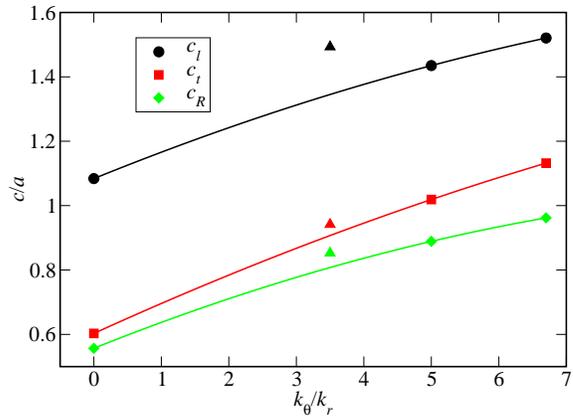}
}
\caption{(Color online) The longitude and the transverse sound wave speeds along with the resulting calculated Rayleigh surface wave speed using Eq. \ref{cr}
for the hcp lattice as a function of $k_{\theta}/k_r$. In circles there are the results with the only 12 $60\degree$ angles that were taken into count, while in the triangles, is
the result with all 24 $60\degree$ angles.}
\label{sounds}
\end{figure}

Since in an hcp lattice, the sound velocities are inhomogeneous, yielding a different sound velocity for each direction, we defined that the relevant variables are the
variables in the XY-plane, which is the major fracture plane, and thus, the crack velocities are normalized to the sound velocities in the XY-plane (which inside
this plane, are homogeneous, as for a 2D hexagonal lattice). In this manner we define the Rayleigh wave speed in the XY-plane, by Eq. \ref{cr}.
We calculate $c_l$ and $c_t$ via measuring the wave velocities by initiating longitude and transverse small deformations in the end of the samples in the different
lattices that we use in this study and then find $c_R$ via Eq. \ref{cr}. The results are shown in Fig. \ref{sounds}, with the  circles indicating  the results
with the only 12 $60\degree$ angles that were taken into account, and the triangles the results with all 24 $60\degree$ angles. The value of $k_{\theta}$ with
all 24 $60\degree$ angles was chosen for reproducing the quantitative values of the model with only 12 $60\degree$ angles model.

We can see that for both lattices, the numerical value for the wave velocities using $k_{\theta}=0$  are very close to the 2D-values~\cite{shay5}, as can be calculated analytically.
For larger values of $k_{\theta}$, the different sounds velocities are higher ($\approx10-15\%$) than the 2D velocities~\cite{shay5}. In addition, we can see that the results
for the sound velocities (in specific, $c_R$) with $k_{\theta}/k_r=3.5$ with all 24 $60\degree$ angles are very much alike the $k_{\theta}/k_r=5$ with only 12 $60\degree$ angles.

\end{document}